# Sub-Natural Linewidth Single Photons from a Quantum Dot

Clemens Matthiesen*, Anthony Nickolas Vamivakas[‡] & Mete Atatüre[*]

*Cavendish Laboratory, University of Cambridge, JJ Thomson Avenue, Cambridge CB3 0HE, UK*

The observation of quantum dot resonance fluorescence enabled a new solid-state approach to generating single photons with a bandwidth almost as narrow as the natural linewidth of a quantum dot transition. Here, we operate in the Heitler regime of resonance fluorescence to generate sub-natural linewidth and high-coherence quantum light from a single quantum dot. The measured single-photon bandwidth exhibits a 30-fold reduction with respect to the radiative linewidth of the QD transition and the single photons exhibit coherence properties inherited from the excitation laser. In contrast, intensity-correlation measurements reveal that this photon source maintains a high degree of antibunching behaviour on the order of the transition lifetime with vanishing two-photon scattering probability. This light source will find immediate applications in quantum cryptography, measurement-based quantum computing and, in particular, deterministic generation of high-fidelity distributed entanglement among independent and even disparate quantum systems.

[*]   cm467@cam.ac.uk and  ma424@cam.ac.uk

[‡]   Present address: Institute of Optics, University of Rochester, 275 Hutchison Rd., Rochester, NY 14627-0186

Numerous proposals in quantum information science ranging from distributed entanglement generation to interfacing disparate qubits rely on single photon generation from state-selective optical transitions[1,2,3,4]. The full quantum description of the resonant interaction of a two-level system with light is captured in the statistical and spectral properties of the scattered light. Shortly after a number of theoretical investigations[5,6,7] quantum effects were observed in the spectrum and photon correlations of atomic-beam resonance fluorescence (RF)[8,9,10]. RF is now routinely employed in optical investigations of atomic gases, single trapped ions and atoms[11]. In the large Rabi frequency limit, the RF spectrum develops into the Mollow triplet[5]. In this regime the generated photons are antibunched on the timescale of the excited state lifetime, they bear no phase relation to the excitation field and their coherence is constrained to the antibunching timescale. Conversely, in the small Rabi frequency limit, two otherwise independent processes of absorption and emission become a single coherent event. In such a process the scattered photon spectrum replicates the excitation field exhibiting comparable first-order optical coherence. The same scattered field also displays strong antibunching in photon correlation measurements. This phenomenon, known as the Heitler effect[12], has been demonstrated in trapped ions using heterodyning[13] and spectral measurement[14,15] techniques.

RF has only recently been observed from solid-state emitters, such as dye molecules[16] and self-assembled quantum dots (QDs)[17,18,19,20]. The strictly resonant excitation was shown to minimize decoherence when compared to nonresonant excitation[21,22,23]. Further, taking advantage of the spin-selective QD transitions, required for spin-based QIS[24], QD RF allowed detailed studies of spin dynamics[25,26], excitation-induced sideband broadening[27] and the realization of single-shot spin readout[28]. Surprisingly, the limit of vanishing Rabi frequency in the resonant interaction has remained relatively unexplored for atoms, ions and any single solid-state emitter[29]. The technical challenge to observe the elastically scattered photons is to suppress sufficiently the laser background and the incoherent component of RF. Here, we demonstrate that, in the limit of the Rabi frequency much smaller than the spontaneous emission rate – the Heitler regime[12], the spectrum and the coherence properties of RF photons are liberated from the QD transition properties and are tailored predominantly by the excitation laser properties.

Our experimental arrangement constitutes a confocal microscope operating at 4 K where the combination of cross-polarization and confocal rejection of the collection fiber suppresses any residual laser by a factor of $10^7$ with respect to the collected RF[30]. Figures 1(a)-(d) present excitation power dependence of the integrated RF spectrum from a QD transition in order to quantify both the collection efficiency of our system and the fraction of residual laser background in the detected photons. In panel (a) the laser frequency is fixed and photodetection events are recorded (open blue circles) as a function of gate bias (the bias Stark-shifts the QD transition). The full-width half-maximum (FWHM) of the Lorentzian curve fit to the data (red curve) is 530 MHz in linear frequency for the excitation power at the saturation point. The inset displays the residual background for large detunings (blue bars) including the contribution of single-photon counting avalanche photodiode (APD) dark counts (yellow bars). In panels (b) and (c) the red and the black triangles display the total photodetection events per second on- and off-resonance, respectively, as a function of excitation power. The solid curves represent the theoretically expected fits for the data in each panel. The dashed blue line indicates the mean dark count level. For an excitation power at the saturation point ($\Omega \sim \Gamma/\sqrt{2}$), where $\Omega$ denotes the Rabi frequency and $\Gamma$ denotes the spontaneous emission rate) we collect approximately $1.25 \times 10^6$ photons per second from RF at the output of our single-mode optical fiber, while the laser and detector background contribute ~0.084% of the total signal. The blue circles in panel (d) show the obtained ratio of the RF signal to total background (SBR) for each laser power along with the excepted power dependence (blue curve). At a tenth of saturation, the residual laser contribution to the background falls below the dark count level and limits the SBR. This degree of RF collection already compares with that obtained from single trapped ions and will allow an increase of electron spin readout fidelity[24] to better than 99% within 30-μsec detection time.

The theoretically expected spectrum of RF in the presence of dephasing when the laser is exactly resonant can be given as[31]

$$S(\Delta v) \propto \frac{1}{2} \frac{\tilde{T}_2^{-1}}{\Delta v^2 + \tilde{T}_2^{-2}} + \frac{1}{2} \frac{1}{\tilde{T}_1 \tilde{T}_2 + \Omega^2} \left( \frac{A\eta/2 - (8\mu)^{-1}(\Delta v - \mu)B}{(\Delta v - \mu)^2 + \eta^2} + \frac{A\eta/2 + (8\mu)^{-1}(\Delta v + \mu)B}{(\Delta v + \mu)^2 + \eta^2} \right)$$
$$+ \frac{1}{\tilde{T}_1 \tilde{T}_2^{-1} + \Omega^2 \tilde{T}_1^2} \delta(\Delta v)$$

(1)

where $\Delta \nu$ is the detuning in linear frequency, $\tilde{T}_1$ and $\tilde{T}_2$ are lifetime and dephasing time, $\Omega$ is the Rabi frequency. $A$, $B$, $\mu$ and $\eta$ are $A = \Omega^2 + \left(\tilde{T}_1^{-1} - \tilde{T}_2^{-1}\right)\tilde{T}_1^{-1}$,

$B = 2\Omega^2 + \left(3\tilde{T}_1^{-1} - \tilde{T}_2^{-1}\right) - 2\left(\tilde{T}_1^{-1} - \tilde{T}_2^{-1}\right)\tilde{T}_1^{-1}$, $\mu = \sqrt{\Omega^2 + \left(\tilde{T}_1^{-1} - \tilde{T}_2^{-1}\right)^2}$ and $\eta = \left(\tilde{T}_1^{-1} + \tilde{T}_2^{-1}\right)/2$.

The first three terms of Eq. (1) originate from the incoherent component while the last term is responsible for the coherently scattered photons. At high excitation power the incoherent component reveals the Mollow triplet, while at low excitation power the spectrum is dominated by the coherent part. Figures 2(a)-(c) display the measured spectrum of quantum-dot RF for three excitation power settings - $\Omega \sim 1.5\Gamma$, $\Omega \sim 0.6\Gamma$, $\Omega \sim 0.22\Gamma$, respectively. Here the laser is fixed on resonance with the QD transition and the RF spectrum is recorded through a scanning Fabry-Perot cavity (FWHM ~29 MHz). The measured spectrum (red circles) in panel (a) displays the expected Mollow triplet together with a sub-natural linewidth peak due to the coherently scattered component along with the theoretically expected lineshape for each component (solid curves). The incoherent part dominates the deconvolved spectrum at this power. In panel (b) both coherent and incoherent components are still visible, but the incoherent component is reduced to a single peak. Figure 2(c) displays the same measurement for a lower excitation power ($\Omega \sim 0.22\Gamma$), where the entire measured spectrum collapses to a cavity resolution-limited sub-natural linewidth Lorentzian. Any residual incoherent component of RF in this regime is less than the detectable level within the spectral resolution and the signal-to-noise ratio of our measurements. The combination of detuning and increased Rabi frequency can be used to further suppresses the fraction of the residual incoherent component for the same level of RF signal[30].

Figures 2(d)-(f) display the intensity-correlation measurements performed on the complete RF emission at the three Rabi frequency settings of Figs. 2(a)-(c) using a Hanbury-Brown and Twiss setup[30]. The raw data for each measurement (red curve) decorates the theoretically expected function (dashed black curve) and its convolution with the measured instrument response function (blue curve). In Panel (d) the onset of Rabi oscillations is visible in the intensity correlation function, which exhibits an oscillatory behaviour in addition to strong antibunching ($g^2(0)=0.01\pm0.02$) after deconvolution). As the Rabi frequency is reduced in the remaining panels the intensity correlation measurement reveals antibunching is sustained fully within a timescale determined by the QD transition lifetime of 760 ps. While the dramatic reduction of the RF bandwidth suggests the wave nature of light is maintained over timescales significantly beyond the excited-state lifetime, the intensity-correlation measurements dictate that the particle-like nature of the photons is preserved even in this regime.

To obtain a more precise value for the photon emission linewidth, we employ field-correlation measurements using a Michelson interferometer[30]. As a lower bound for single-photon coherence, the red squares in Fig. 3(a) display the interference fringe visibility for QD emission under low power incoherent excitation. The resulting coherence time is ~540 ps, which is substantially shorter than 1.52 ns – the theoretical upper limit of $T_2=2T_1$ for the investigated QD – and is consistent with recent reports on modest degree of two-photon interference from independent QDs[31,32]. The dashed black line in Fig. 3(a) represents the measured fringe visibility for the laser and limits the maximum attainable coherence time in our experiments to ~50 ns. The blue squares display fringe visibility for RF generated under the same condition as for Figs. 2(c) and 2(f). For the coherently scattered single photons the measured coherence time is significantly longer than the theoretical limit for spontaneously emitted photons, where absorption and emission can be considered as two separate physical events. The theoretical fit (blue curve) reveals a coherence time of 22 ns, which corresponds to a spectral width upper bound of 7 MHz indicating at least a 30-fold reduction of the single photon bandwidth with respect to the transition linewidth. Second, the theoretical fit to the measurement at short and long temporal delay regimes reveals that 9% of the total scattering constitutes the incoherent component with a coherence time comparable to $2T_1$ within the measurement uncertainty. As expected, this incoherent fraction further reduces to only 5% at lower excitation power ($\Omega\sim0.17\Gamma$). In this regime the interference fringe visibility of RF (green circles) after a delay of ~2.6 ns is as high as 85% of that displayed by the coherent laser. These results reveal the counter-intuitive nature of RF in the Heitler regime: The QD operates as a single-photon turnstile device acting on a coherent beam. The photons arrive one at a time to our detection system, but are phase-correlated with each other and with the laser field from which they originated. In this regime, issues thought to be inherent in solid-state systems, such as spectral diffusion, modify the probability to scatter a photon, but do not cause photon decoherence due to the absence of an exciton interacting with the environment.

In this work we have demonstrated that a solid-state quantum emitter such as a single QD can generate single photons with bandwidth and coherence that are tailored by the excitation laser. Consequently, an immediate opportunity emerges where tunable and time-synchronized single photons can be generated from multiple QDs with near-arbitrarily tailored spectral and phase properties using the same excitation laser. In contrast to using an ultrafast pulse to trigger an excited state population[33,34] for single photons, our approach lends itself to the use of nanosecond pulses where there is significant technical ease in controlling the amplitude and phase of the excitation field. Such custom-designed photonic states will facilitate high-fidelity photonic coupling between otherwise disparate quantum systems, such as spectrally distinct QDs and trapped ions[4], as required for the realization of distributed hybrid quantum networks[1].

This work was supported by grants and funds from the University of Cambridge, EPSRC Science and Innovation Awards, the QIPIRC and EPSRC grant number EP/G000883/1. We thank C. Stroud, Jr., M. Kohl and Z. Hadzibabic for useful discussions, and P. Humphreys and C.-Y. Lu for technical assistance. We further thank M. Hughues, M. Hopkinson and the National Centre for III-V Technologies for providing the QD sample used in this work.

**Figure Captions**

**FIG. 1.** Collection efficiency and background analysis of resonance fluorescence. **(a)** Integrated photon counts for RF as a function of laser detuning. The red curve is a Lorentzian fit to the data (open circles). Inset: Total off-resonance (~35 GHz) background counts (blue data) compared to the APD dark counts alone (yellow data). **(b)** Integrated photon detection events when the laser is on resonance with the QD transition (red triangles) and **(c)** off resonance (black triangles). **(d)** RF signal to background ratio (SBR). At saturation ($\Omega \sim \Gamma/\sqrt{2}$) a value of 1050 is achieved.

**FIG. 2.** Spectrum and intensity-correlation measurements on QD RF. **(a)** A narrowband single mode laser is fixed on resonance with the QD transition at $\Omega \sim 1.5\Gamma$ and the RF spectrum is recorded by an APD as the transmission through a scanning Fabry-Perot cavity (FWHM ~29 MHz), shown as red circles. The spectrum comprises the onset of the Mollow triplet (fitted with the dark green curve) superimposed by the coherently scattered component (fitted with the light green curve). The grey circles display the residual laser background under the same conditions, but when the QD transition is gate-detuned. **(b)** Same measurement as panel (a) performed when the laser power is around saturation ($\Omega \sim 0.6\Gamma$) and **(c)** when the laser power is below saturation ($\Omega \sim 0.2\Gamma$). The spectrum collapses to a single Lorentzian of width indistinguishable from the resolution of the scanning cavity. **(d-f)** Intensity correlation measurements of RF for the excitation power regimes of the spectra presented in panels (a-c), respectively. In all the plots the data (red curves) are superimposed by the theoretical prediction (blue curves) convolved with the system response function[29], while the dotted black lines indicate how our measurements would appear for an ideal system response. All plots display the ubiquitous antibunching behavior at zero time delay.

**FIG. 3.** First-order correlation measurement of QD RF. **(a)** The visibility of the interference fringes observed for RF for the laser excitation power corresponding to $\Omega \sim 0.22\Gamma$ (blue squares) and $\Omega \sim 0.17\Gamma$ (green squares). The black dashed line is the measured mean fringe visibility for the laser and marks the highest attainable coherence of the scattered photons. The first-order correlation measurement for incoherently generated photons (above GaAs bandgap excitation) (red squares) shows a coherence time of ~540-ps, while the RF displays a coherence time of ~22

ns in the Heitler limit, i.e. ~30 times longer than the antibunching timescale. The detailed description of the theoretical fits to the data (blue, green and red solid curves) can be found in the SOM. **(b)** A close-up of the raw data for RF displaying the observed fringes for a relative time delay of 330 ps. **(c)** Same as panel (b) for a relative time delay of 2.68 ns.

Figure 1.

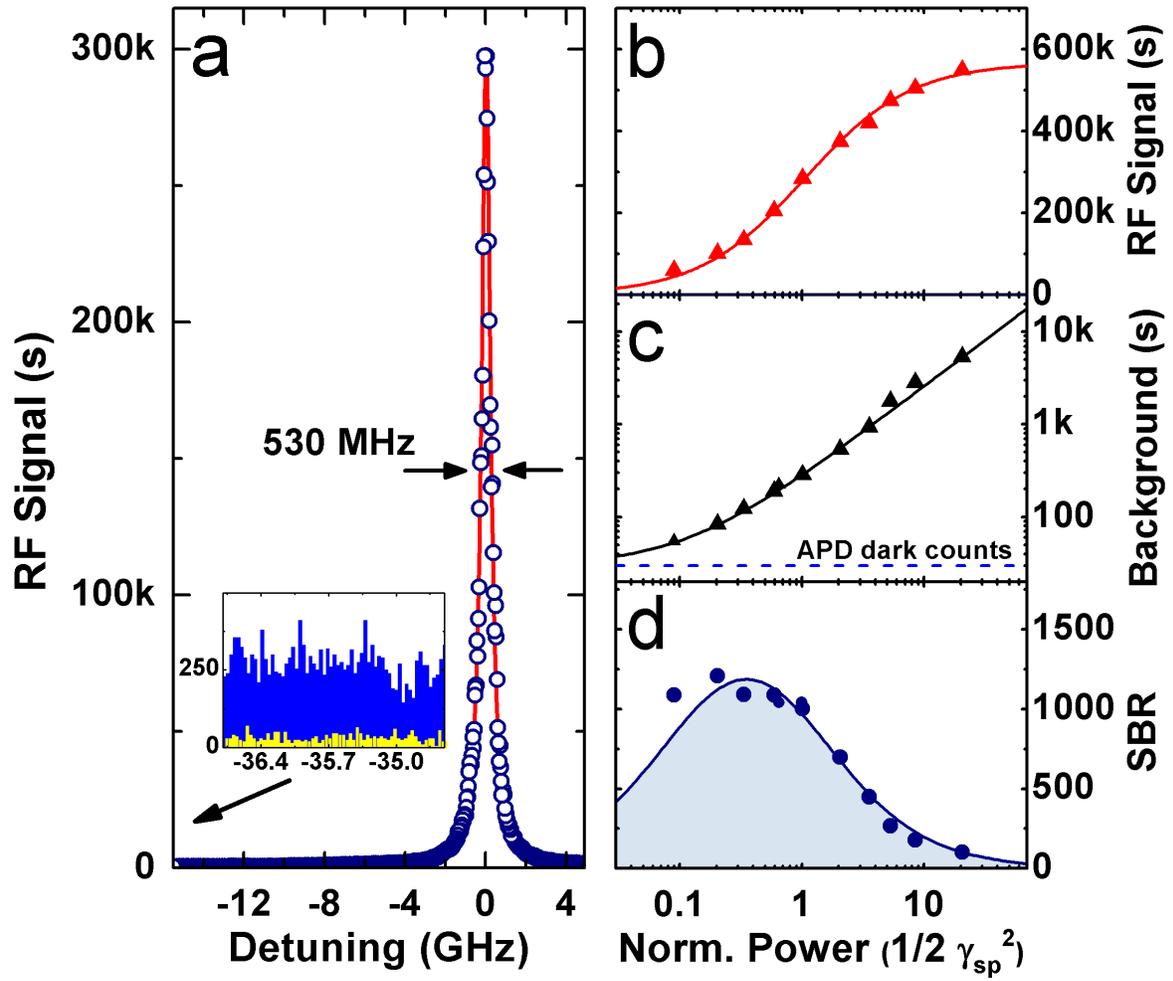

**Figure 2.**

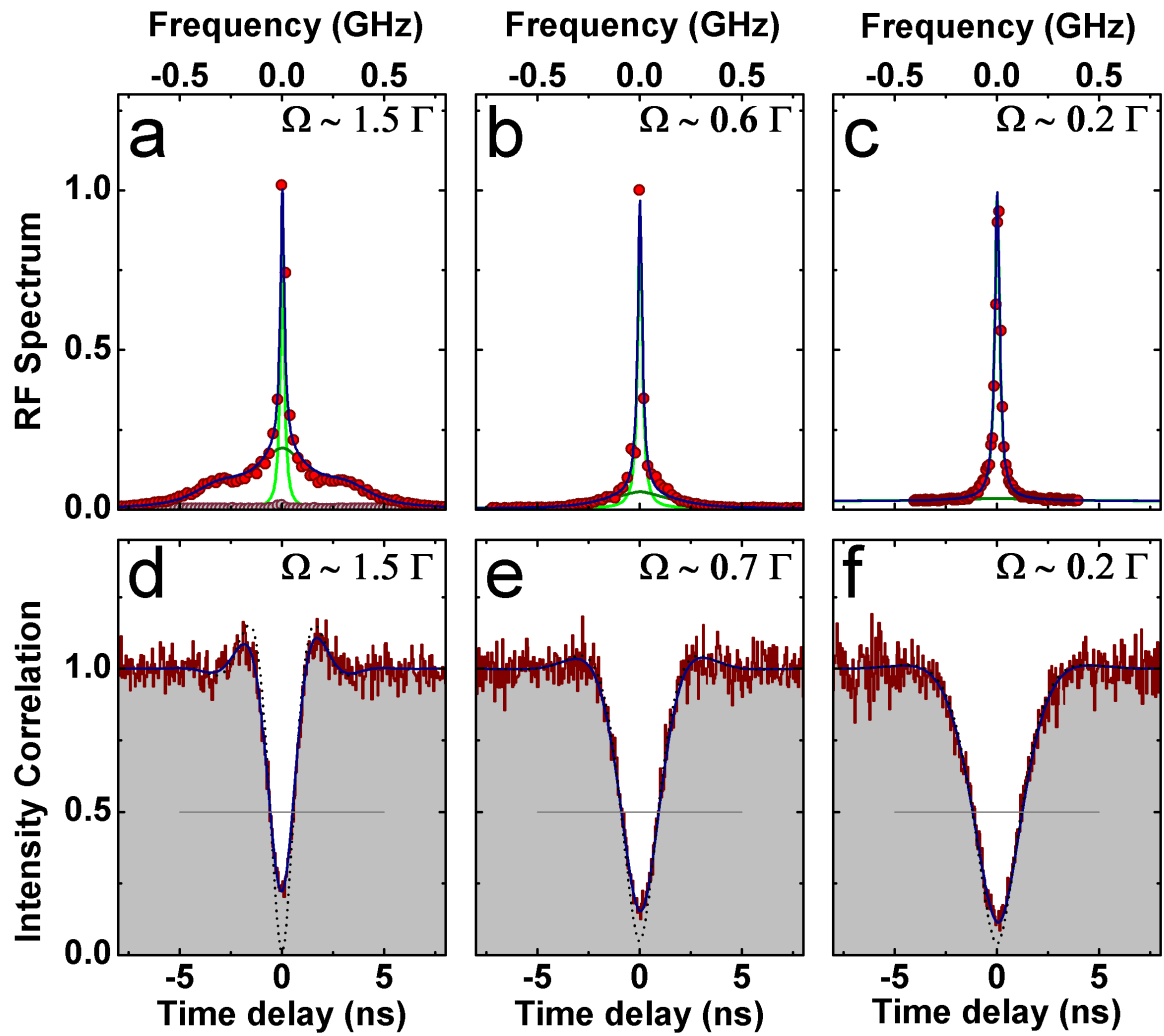

**Figure 3.**

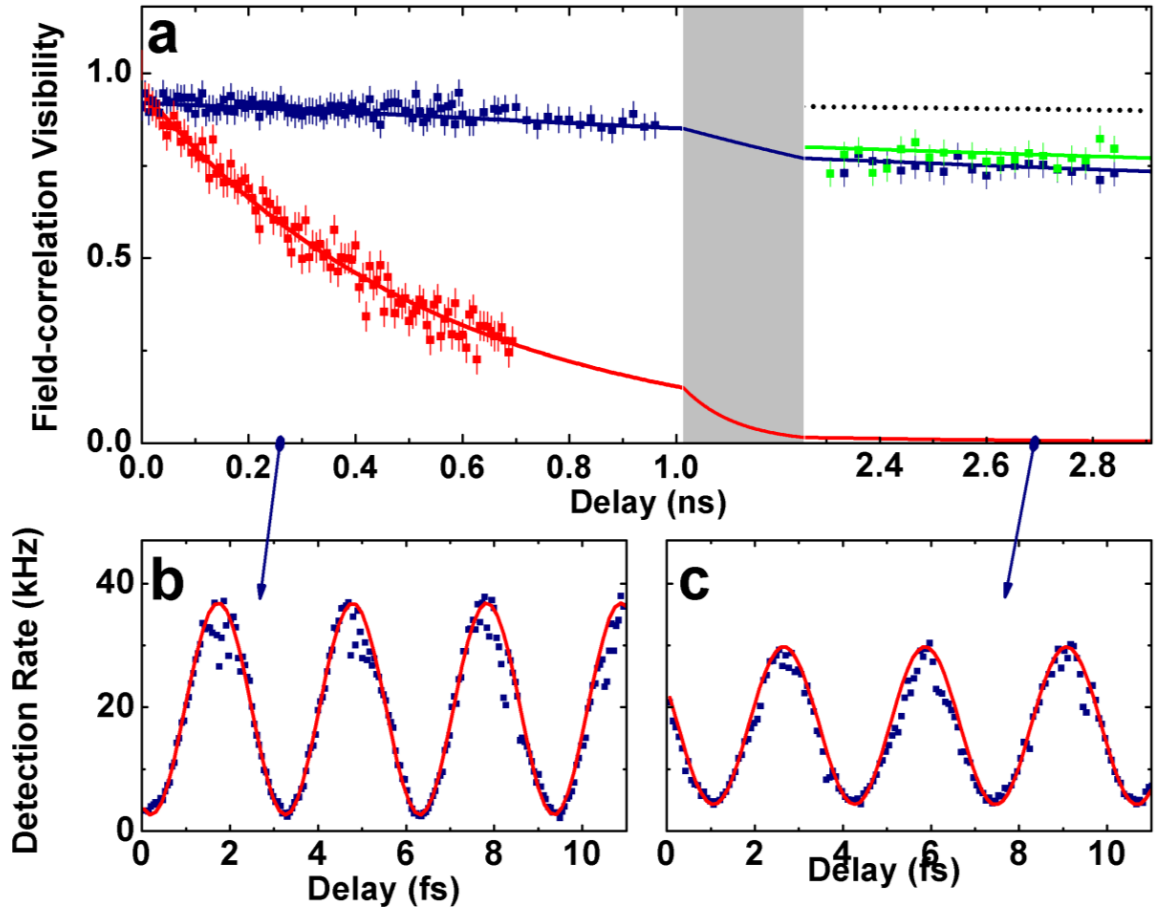

# Supplementary Information:

# Sub-Natural Linewidth Single Photons from a Quantum Dot

Clemens Matthiesen, Anthony Nickolas Vamivakas & Mete Atatüre

*Cavendish Laboratory, University of Cambridge, JJ Thomson Avenue, Cambridge CB3 0HE, UK*

This document contains supplementary data, description and analysis for the results and discussions presented in the manuscript. It includes:

I. The description of the sample and the experimental setup.
II. The effect of spectral detuning on the fraction of the incoherent part of resonance fluorescence.
III. The description of the theoretical fits to the spectrum, the field-correlation and the intensity-correlation measurements including measured response functions.

## I. The sample and the experimental setup:

The sample investigated in this work is Gallium Arsenide with a single layer of self-assembled Indium Arsenide quantum dots (QDs) grown by molecular beam epitaxy. It includes, from bottom to top, a distributed Bragg reflector (DBR) consisting of 20 layers of GaAs/AlGaAs such that reflection in the spectral region 960-980nm is enhanced, an n+-doped layer, QD and capping layers. A Schottky diode heterostructure is obtained by gating the n+-layer with a 5-6 nm thick titanium layer evaporated on top of the sample surface. This structure allows for deterministic charging of the QDs and shifting of the QD exciton energy levels via the DC Stark effect [1]. The relative position of the QD layer with respect to the DBR places the QDs in an antinode of the reflected field revealing ~2-fold photon-collection enhancement with respect to a non-DBR sample. A Zirconia super-solid immersion lens glued onto the sample surface increases the effective numerical aperture of the imaging system. The sample is situated in a bath cryostat at 4.2K. The QD laser excitation and fluorescence collection are achieved with a confocal microscope. A 0.5 NA objective lens is used to excite single QDs and collect QD fluorescence and the laser reflection along the same path into a single mode fiber. A 10/90 beamsplitter separates excitation from collection. The laser is linearly polarized and a polarizer in the collection arm is manually rotated to extinguish laser reflection. Cross-polarisation and confocal rejection of the collection fiber suppresses any residual laser by a factor of $10^7$. Care is taken to couple the QD fluorescence efficiently into the single-mode fiber but not the residual laser reflection.

The collected fluorescence is sent to a free-space Michelson interferometer for field-correlation measurements. Short- and long-delay measurements shown in Fig. 3 in the manuscript are performed by introducing two sets of fixed path delay to the scanning arm of the interferometer. One of the interferometer outputs is coupled into a single mode fiber coupled directly to an avalanche photodiode (APD) to record single photon interference. In addition to the interference signal, the photon counts per interferometer arm (by blocking one arm, then the other) as well as overall dark counts (by blocking both arms) and residual laser leakage (by taking the QD off resonance) is also recorded as a periodic sequence for calibration and reference.  Intensity-correlation measurements are performed by sending the collected signal to a Hanbury-Brown and Twiss (HBT) setup comprising a beam splitter, two APDs, a time-to-amplitude converter and a peak-height analyser. The system response is quantified by using a pulsed (~2 ps) laser at comparable photodetection rate as the corresponding signal. The spectrum of resonance fluorescence is measured through a laser-stabilized scanning Fabry-Perot cavity with a resolution of 29 MHz, measured by scanning the cavity across the excitation laser (see Fig. S2 A).

## II. Resonance fluorescence spectrum with spectral detuning:

The measurements presented in Fig. 2(a)-(c) in the manuscript are obtained for the resonance condition. Figure S1 presents the resonance fluorescence spectrum measured by the scanning Fabry-Perot cavity for a range of QD-laser spectral detuning values. Here, the QD resonance is Stark-shifted by 220MHz for each scan. At fixed laser power the strongest overall signal is observed for the resonance condition (0,0) and the incoherent scattering fraction is highest. As the QD-laser detuning is increased, the overall signal reduces, as well as the fraction of incoherent scattering with respect to the total fluorescence, as expected.

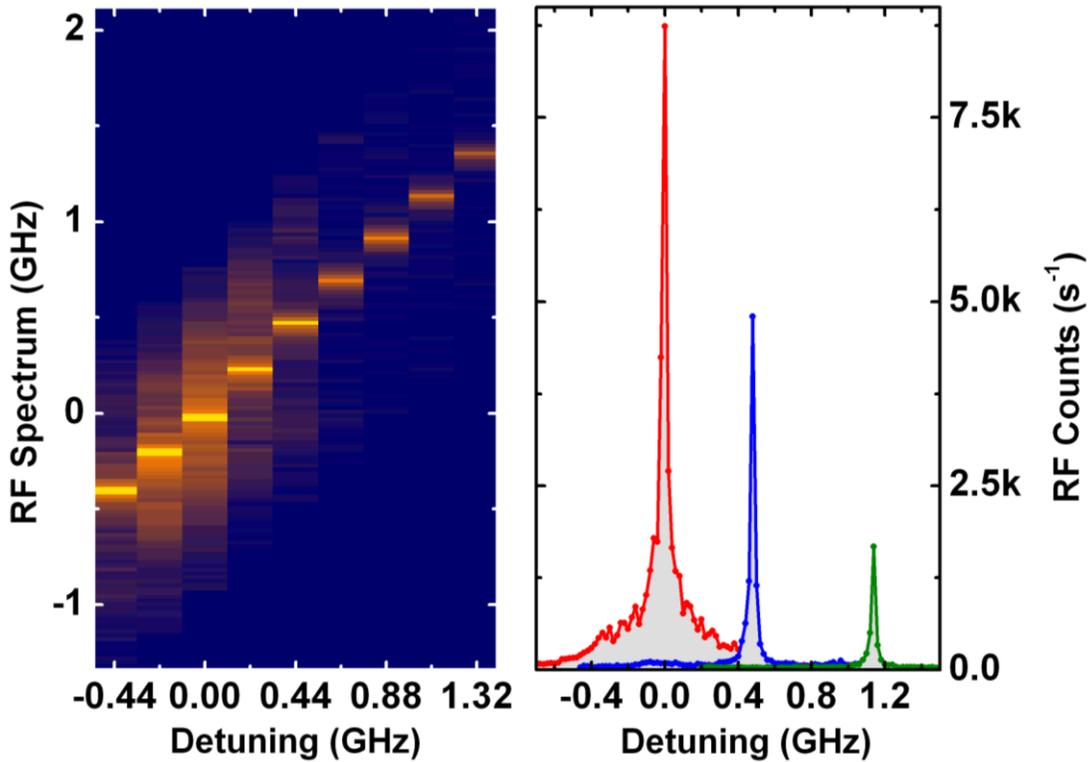

**FIG. S1.** Resonance fluorescence spectra for a set of detuning values applied to the QD transition. The Rabi frequency on resonance corresponds to $\Omega \sim \Gamma$. The sharp peak is the coherently scattered part while the broad feature on resonance is the incoherent part of resonance fluorescence. The residual laser is negligible (not observable) in this measurement.

## III. Theoretical fits to the spectrum and the correlation measurements:

The full *spectrum of resonance fluorescence* from a QD transition driven resonantly in the presence of pure dephasing [2] can be decomposed into incoherent and coherent parts. The incoherent part is given by

$$S_{inc}(\Delta v) \propto \frac{1}{2}\frac{\tilde{T}_2^{-1}}{\Delta v^2 + \tilde{T}_2^{-2}} + \frac{1}{2}\frac{1}{\tilde{T}_1\tilde{T}_2 + \Omega^2}\left(\frac{A\eta/2 - (8\mu)^{-1}(\Delta v - \mu)B}{(\Delta v - \mu)^2 + \eta^2} + \frac{A\eta/2 + (8\mu)^{-1}(\Delta v + \mu)B}{(\Delta v + \mu)^2 + \eta^2}\right)($$

1)

where $\Delta v$ is the detuning in linear frequency, $\tilde{T}_1$ and $\tilde{T}_2$ are lifetime and dephasing time in angular time, $\Omega$ is the Rabi frequency. $A$, $B$, $\mu$, $\eta$ are shorthand for

$$A = \Omega^2 + \frac{\tilde{T}_1^{-1} - \tilde{T}_2^{-1}}{\tilde{T}_1}, \quad B = 2\Omega^2 + (3\tilde{T}_1^{-1} - \tilde{T}_2^{-1}) - 2\frac{(\tilde{T}_1^{-1} - \tilde{T}_2^{-1})^2}{\tilde{T}_1}, \quad \mu = \sqrt{\Omega^2 + (\tilde{T}_1^{-1} - \tilde{T}_2^{-1})^2} \quad \text{and}$$

$\eta = \left(\tilde{T}_1^{-1} + \tilde{T}_2^{-1}\right)/2$. The coherent scattering reflects the coherence of the excitation field and is usually described as a delta function $\delta(\Delta v)$ on resonance While the elastic scattering can be approximated as a delta function for our spectral measurements due to limited resolution, we have to bear in mind that the laser itself is of finite coherence, therefore a more precise description of this term would include the delta function integrated over the laser bandwidth. Further, we cannot exclude that other mechanisms can broaden elastic scattering beyond this upper bound. Its relative magnitude compared to the total scattering is [3]

$$\frac{I_{coh}}{I_{coh} + I_{inc}} = \left(1/2 + T_1/T_2 + 2\Omega^2 T_1^2\right)^{-1}. \tag{2}$$

RF spectra were measured on resonance and for a set of fixed detunings for 7 different powers between $\Omega \sim 0.22\Gamma$ and $\Omega \sim 1.5\Gamma$. The measurements are fitted using (scaled) convolutions of the expected behavior from (1) and (2) with fitted response function. An exemplary response function obtained by scanning the FP cavity across the excitation laser is shown in Fig. S2 A. In the fits $T_2$ and the scaling factor are varied while the values for $T_1$ (=760 ps) and $\Omega$ are fixed. $T_1$ is determined through $g^{(2)}$ measurements and the Rabi frequency can be related to the measured excitation power P as $\Omega^2 = \Omega_{sat} \frac{P}{P_{sat}} = \frac{P/P_0}{T_1 T_{2,sat}}$ where $P_{sat}$ ($\Omega_{sat}$) is the saturation power (Rabi frequency). The dephasing times extracted by fitting are $2T_1$ limited for the lowest Rabi frequency ($\Omega \sim 0.22\Gamma$), decreasing to ~0.7 of this theoretical maximum at the highest Rabi frequency shown in the manuscript.

The theoretical visibility of the *field-correlation* measurement performed on QD resonance fluorescence follows the Fourier transform of the spectrum given above. The delta function in frequency of the elastic scattering gives rise to a constant contribution in the visibility, according to (2). This assumes a source of infinite coherence and no further dephasing effects, however. To adjust for experimental imperfections we give the coherent contribution a finite width (and corresponding coherence time), where the coherence time is a fit parameter. The laser coherence (width of ~3 MHz) of ~53 ns marks the upper limit of the field-coherence used in our analysis.

The incoherent component is expected to follow [2]:

$$G^{(1)}(\tau) \propto 1/2\exp-(\tau/T_2) + \exp-(\tau/2(T_2^{-1} + T_1^{-1}))[N\cos(\Omega'\tau) + M\sin(\Omega'\tau)]. \quad (3)$$

Here, $\tau$ is the delay between the interfering field components, $\Omega' = \sqrt{\Omega - (T_1^{-1} - T_2^{-1})^2/4}$,

$N = \frac{1}{2}\frac{\Omega^2 - (T_1^{-1} - T_2^{-1})/T_1}{\Omega^2 + (T_1 T_2)^{-1}}$ and $M = -\frac{1}{4\Omega'}\frac{\Omega^2(T_2^{-1} - 3T_1^{-1}) + (T_1 - T_2)^2/(T_1^3 T_2^2)}{\Omega^2 + (T_1 T_2)^{-1}}$. The fit parameters are $T_2$, a scaling factor and the coherence (and hence the linewidth) of the coherent part, while $T_1$ (=760 ps) and $\Omega$ are kept fixed. For the results shown in the main text (low Rabi frequency regime) we obtain $T_2 \sim 2T_1$ ($T_2 = 1.52 \pm 0.15$ ns for $\Omega \sim 0.17\Gamma$ $T_2 = 1.45 \pm 0.15$ ns for $\Omega \sim 0.22\Gamma$) indicating that pure dephasing of the incoherent component is essentially suppressed when the spectrum is dominated by the coherent part, consistent with the spectral measurements. The coherence of the elastic component is determined to be $(22\pm 3)$ns (or ~7MHz linewidth).

The *intensity autocorrelation* for a resonantly driven single photon emitter is captured by

$$g^{(2)}(\tau) = 1 - \exp(-\eta|\tau|)[\cos(\mu|\tau|) + \eta/\mu \sin(\mu|\tau|)], \tag{4}$$

with parameters defined as above. Data were taken in the range of $\Omega \sim 0.22\Gamma$ and $\Omega \sim 3.5\Gamma$ and fit by expression (4) convolved with a measured system response function, displayed in Fig. S2 B. $T_2$ is the sole fitting factor, $T_1$ and $\Omega$ are fixed as in all other fits. Again, consistent with spectra and field-correlation we observe a vanishing of pure dephasing at low Rabi frequencies.

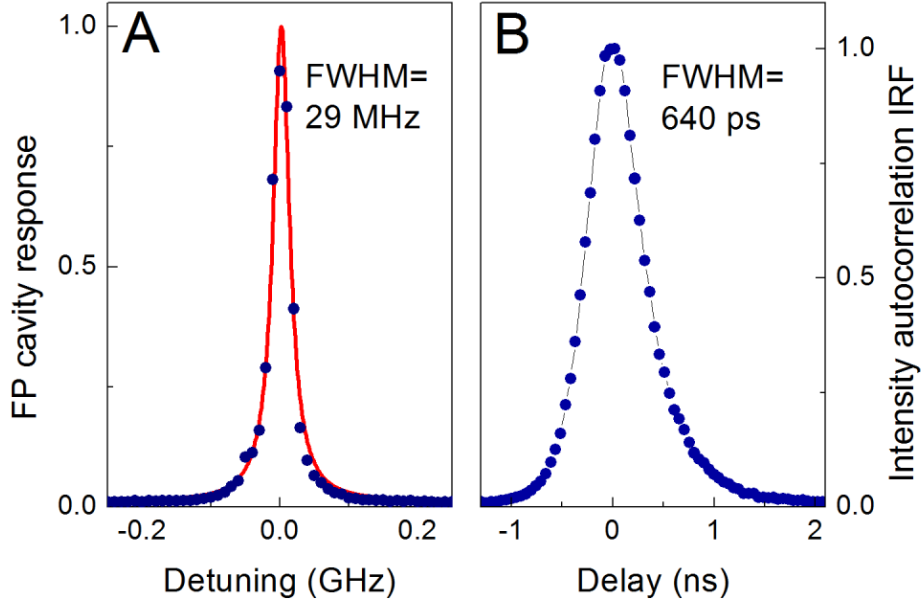

**FIG. S2.** Instrument response functions (IRF) for spectral measurements (a) and intensity autocorrelations (b). **(a)** A Lorentzian fit of full-width at half-maximum (FWHM) of 29MHz is shown as red line. **(b)** Response of the intensity autocorrelation measurement system to ~2ps laser pulses.

It is interesting to note that while the spectrum of RF in the Heitler regime collapses to that of the excitation laser yielding significantly longer photon coherence times in comparison to the characteristic lifetime of the QD transition, the photon-correlation measurements still reveal that only one photon is scattered within the timescale set by $1/\mu$, where $\mu$ is defined in Eq. (1). At low power this timescale approaches the excited-state lifetime, while in the high excitation power regime it is reduced and eventually dictated by $\Omega^{-1}$. Therefore, if a finite bandwidth is enforced on the coherent component by an optical filter (such as a cavity) or limited bandwidth of detection apparatus, the essential information on the photon arrival times (hence the antibunching signature) can be erased to eventually recover a Poissonian distribution for photon number at a given time. It is nevertheless interesting that in the spectrum picture, the imposed filter bandwidth that erases the antibunched nature of RF can still be orders of magnitude broader than the photon linewidth.

The measurement techniques reported here (spectral, field- and intensity-correlation measurements), independently use $T_2$ for fitting and allow us to observe a dephasing dependence on Rabi frequency. Unlike the $\sim\Omega^2$–dependence of the photon dephasing rate reported for a much higher excitation-power regime [4] for a continuous wave laser, or even higher excitation power regime due to strong pulsed excitation [5], we only observe a linear ($\sim\Omega$) dependence as extracted from a total of 19 separate but consistent fits to full spectrum, field- and intensity-correlation measurement covering the excitation power regime reported in this work, where every other parameter is kept constant. This dependence suggests an alternative dephasing mechanism to be dominant in the power regime discussed here and will be considered in a future study.